\documentclass[journal=aanmf6,manuscript=article,email=false]{achemso}

\graphicspath{{Figures/}}
\usepackage{chemformula} 
\usepackage[T1]{fontenc} 
\usepackage{setspace}

\usepackage{tikz,scalerel,MnSymbol}

\newcommand{\tikzcircle}[1][red,fill=red]{\scalerel*{\tikz \draw[#1] (0,0) circle (1.5pt);}{\square}}

\newcommand{\tikzsquare}[1][red,fill=red]{\scalerel*{\tikz \draw[rounded corners=0.06pt,#1] (0,0)--++(90:3pt)--++(0:3pt)--++(-90:3pt)--cycle;}{\square}}

\author{Movaffaq Kateb}
\affiliation{Condensed Matter and Materials Theory Division, Department of Physics, Chalmers University of Technology, SE-412 96 Gothenburg, Sweden}

\author{Andrei Manolescu}
\affiliation{Department of Engineering, School of Technology, Reykjavik University, Menntavegi 1, IS-102 Reykjavik, Iceland}

\title[Partially oxidized nanosolids]
  {Distinct melting behaviour of partially oxidized Cu nanoparticles and nanowires}


\begin{document}


\begin{abstract}
\begin{singlespace}
While most of the experimental studies are dealing with partially oxidized metallic nanosolids, their behaviour is not well understood theoretically. To this end we utilized molecular dynamics simulation and charge-optimized many-body potential to probe melting of Cu nanoparticles and nanowires with and without a monolayer of Cu$_2$O as a shell. It is shown that partially oxidized nanosolids present a different melting behaviour than the widely accepted picture of melting, and especially different than surface melting. (i) For both types of nanosolids we observed inward diffusion of $\sim$40\% of atomic oxygen into the metallic core at room temperature. (ii) For the nanowire with oxide shell we observed a solid state phase transition in the Cu core  that is not present for the case without the oxide. (iii) Prior to melting, the oxide shell shrinks, and this process continues after melting, to form a particulate oxide, for both types of nanosolids.
\end{singlespace}
\end{abstract}

\begin{singlespace}

\section{Introduction}
Metallic nanosolids have proven promising in various applications due to their electronic and optical properties, extraordinary catalytic behaviour, or biological activity. \citep{poole2003} Since the beginning, understanding their thermal stability has been of both fundamental interest and practical importance. \citep{michaud2019} In this regard, the melting temperature of nanosolids, $T_{mn}$, has received a considerable attention. \citep{gao2016} This is partially due to fact that $T_{mn}$ can be modeled assuming the empirical relation between the bulk melting temperature, $T_{mb}$, and cohesive energy, $E_b$, stands true at the nanoscale. \citep{goldstein1992,jiang1999,safaei2010} Along with experimental results, these models were able to explain phenomenological properties of metallic nanoparticles that could not be observed in the bulk. In particular, individual nanoparticles have $T_{mn}$ well below $T_{mb}$ (\textit{depression}) \citep{safaei2010} while embedded ones might have higher $T_{mn}$ than the corresponding bulk material (\textit{superheating}). \citep{sheng1996} However, since the models were based on the ground state crystal structure and nanosolids shape, they failed to describe solid state phase changes and melting dynamic or atomistic mechanisms. \citep{kateb2018,azadeh2019}

On the other hand, atomistic simulations such as molecular dynamics (MD) and Monte Carlo (MC), have proven to be excellent tools for probing the nanosolids behavior at elevated temperatures. \citep{gao2016} For instance, the size-dependent $T_{mn}$ of spherical Cu nanoparticles  \citep{wang2003,delogu2005,kart2014,loulijat2015} and cylindrical Cu nanowires \citep{zhang2010nw, zhang2018} has been observed using MD simulation. In this regard, the major effort has been devoted to demonstrate size dependent surface melting below $T_{mn}$. Recently, it has been shown that both naoparticles \citep{liang2017} and nanowires \citep{beloshapka2019} perform size-dependent reshaping prior to melting. More recently, we showed that Pd nanoparticles present size-dependent solid-state transitions below the melting point. \citep{kateb2018,azadeh2019} The latter has been also verified for the Cu nanorods. \citep{zhang2018} In relation with $T_{mn}$ depression, the coalescence and densification of two or more spherical Cu particles have also been studied by means of MD simulation. \citep{li2017,yang2018,tsai2019} While most of these simulations consider an ideal crystallite, in practice Cu nanoparticles suffer from the presence of defects, contamination, surfactants/capping, or partial oxidation. \citep{yeshchenko2007,jeong2008} Among these, the surface oxide has been widely studied experimentally, as Cu nanoparticles can be easily oxidized, e.g.\ upon exposing to air or during wet chemical synthesis. \citep{nakamura2007,chernavskii2007,pena2011,tan2013,leitner2020} Thus, including an oxide shell is a natural maturing step towards a more realistic simulation. 

Surprisingly, the effect of an oxide shell on the melting behaviour of nanoparticles using MD simulation has been barely studied. \citep{puri2010} \citet{puri2010} studied spherical Al nanoparticles of different diameters  including crystalline and amorphous oxide shells with variable thicknesses. Most importantly, they utilized \citet{streitz1994} charge density distribution, which leads to a smooth variation of the potential function at the metal-oxide interface, while a point charge assumption can result in a discontinuity in the potential. Thus, they have been able to verify the outward diffusion of the metallic core that was previously reported in experimental studies. \citep{nakamura2007} However, they utilized embedded atom method (EAM) as their short ranged interaction, that does not include the bond order, and thus cannot properly describe covalent bonding nature of metal oxides. Nevertheless, they calculated mean square displacement instead of \citet{lindemann1910} index ($\delta_{\rm L}$). The former is proportional to the diffusion coefficient while the latter reflects an average over the amplitude of atomic vibrations. Thus, they have only been able to visually determine $T_{mn}$.

In the present work, the melting behaviour of Cu nanoparticle and nanowire is investigated with and without an oxide shell. We choose Cu as the most popular nanoink for printing interconnects on flexible substrates. \citep{gao2016} The 5~nm in diameter is considered as it has been shown promising in post-printing annealing. \citep{kim2009} Besides, Cu oxidation at the nanoscale is well established and its oxide thickness can be carefully controlled. \citep{jeong2008} It is worth mentioning that, unlike Al oxides, Cu oxides are not stable over a wide range of temperatures, while metallic Cu has higher $T_{mb}$ than Al. Thus, one can expect different results than the earlier study of \citet{puri2010}. To this end, the variable charge scheme is utilized within the MD framework. Unlike \citet{puri2010}, we utilize \citet{tersoff1988} potential as a short range interaction that can properly capture the bond order aspect of the oxide. Furthermore, various melting criteria are utilized to ensure a comprehensive picture of the melting process.

\section{Method}
MD simulations were performed by solving Newton's equation of motion \citep{allen1989} using LAMMPS open source code\footnote[1]{Version~7~Aug.~2019}. \citep{plimpton1995,plimpton2012} We utilized charge-optimized many-body (COMB) potential \citep{devine2011} in which the COMB notation, the total energy of system is expressed as
\begin{equation}
    U^{\rm Tot}(q,\mathbf{r})=U^{\rm ES}(q,\mathbf{r})+U^{\rm Short}(q,\mathbf{r})+U^{\rm Corr}(\mathbf{r}) \ ,
\end{equation}
where the terms on the right respectively are electrostatic, short-ranged and correction terms, with $q$ and $\mathbf{r}$ are being the charge and coordinates of atoms, respectively. The $U^{\rm ES}$ itself consists of self energy ($U^{\rm Self}$) and energies due to charge densities ($U^{qq}$) and charge-nuclear ($U^{qZ}$) contributions,
\begin{equation}
    U^{\rm ES}(q,\mathbf{r})=\sum_iU_i^{\rm Slef}(q)+\frac{1}{2}\sum_{i\neq j}q_iJ_{ij}^{qq}q_j+\sum_{i\neq j}q_iJ_{ij}^{qZ}q_j \ .
\end{equation}
The self energy  $U_i^{\rm Self}$ of atom $i$ can be calculated using a simplified  expression based on a Taylor series,
\begin{equation}
    U_i^{\rm Self}(q,\mathbf{r})=\chi_iq_i+(J_i+J_i^{\rm field})q_i^2+K_iq_i^3+L_iq_i^4 \ ,
\end{equation}
and the Coulomb integrals for charge densities ($\rho_{i,j}$) are given by  \ ,
\begin{equation}
    J_{ij}^{qq}
    =\int d^3x_i\int d^3x_j\frac{\rho_i(\mathbf{x}_i,\mathbf{r}_i)\rho_j(\mathbf{x}_j,\mathbf{r}_j)}{x_{ij}} \  ,
\end{equation}
where $\mathbf{r}_{i}$ and $\mathbf{r}_{j}$ are the positions of the nuclei,  
and $x_{ij}$ is distance between the spatial points points $\mathbf{x}_i$ and 
$\mathbf{x}_j$. Finally, the charge-nuclear coupling operator is equal to
\begin{equation}
    J_{ij}^{qZ} 
    =\int d^3 x\frac{\rho_i(\mathbf{x},\mathbf{r})}{|\mathbf{x}-\mathbf{r}_i|} \ ,
\end{equation}
and $\rho_i$ can be described by \citet{streitz1994} rigid density distribution
\begin{equation}
    \rho_i(\mathbf{r},q_i)=Z_i\delta_{\rm Kron}(|\mathbf{r}-\mathbf{r}_i|)+ (q_i-Z_i)\frac{\xi_i^3}{\pi}\exp\big(-2\xi_i|\mathbf{r}-\mathbf{r}_i|\big) \ .
\end{equation}
Here $Z_i$ is atomic number, $\delta_{\rm Kron}$ is Kronecker delta function and $\xi$ is the orbital exponent that determines decay with respect to $\mathbf{r}_i$.

The $U^{\rm Short}$ term combines attractive and repulsive terms of general Tersoff potential
\begin{equation}
    U^{\rm Short}(r_{ij},q_i,q_j)=f_c(r_{ij})\big[A_{ij}\exp(-\lambda_{ij}r_{ij})-b_{ij}B_{ij}\exp(-\alpha_{ij}r_{ij})\big] \ ,
	\label{eq:short}
\end{equation}
with $A_{ij}$, $B_{ij}$, $\lambda_{ij}$ and $\alpha_{ij}$ being fitting parameters and $f_c$ being the smoothing function that works near the cutoffs. The main bond order term of Tersoff potential is $b_{ij}$ that changes the attraction based on the bond angle, number of nearest neighbors, and their symmetry:
\begin{equation}
    b_{ij}=\big[1+(\beta\zeta_{ij})^n\big]^{-\frac{1}{2n}} \ ,
\end{equation}
\begin{equation}
    \zeta_{ij}=\sum f_c(r_{ij})g(\theta_{ijk})\exp\big[\lambda^m(r_{ij}-r_{ik})^m] \ ,
\end{equation}
\begin{equation}
    g(\theta_{ijk})=\Big(1+\big(\frac{c}{d}\big)^2-\frac{c^2}{d^2+(h-\cos\theta_{ijk})^2}\Big)\gamma_{ijk} \ ,
\end{equation}
where $\alpha$, $\beta$, $n$, $m$, $c$, $d$, and $h$ are fitting constants. It is worth mentioning that this is the general formalism of Tersoff potential and some of these parameters are known constants depending on the specific formalism. The term $U^{\rm Corr}$ also includes some charge-dependent correction into $B_{ij}$ and exponential terms of Eq.~(\ref{eq:short}).

Furthermore, COMB potential requires the charge equilibration (QEq) introduced by \citet{rappe1991} over the time span of simulation. Briefly, QEq determines the charge distribution based on the atomic ionization energy, electron affinity and geometry to minimize the electrostatic energy. The, charge and coordinates of atoms can be linked through the Lagrangian \citep{rick1994}
\begin{equation}
    L=\sum_i\frac{1}{2}m_i\dot{r}_i^2+\sum_i\frac{1}{2}M_q\dot{q}_i^2-U^{\rm Tot}(q,\mathbf{r})-v\sum_iq_i \ ,
\end{equation}
where $m_i$, $r_i$ and $q_i$ are atomic mass, coordinates and charge, respectively. $M_q$ is a fictitious charge mass and $v$ is the tolerance of the charge conservation.

It is worth mentioning that the COMB potential overestimates the Cu melting temperature. We justify using COMB for three reasons. First, here the main focus is on the oxide shell and COMB can carefully model the Cu-oxide interface interactions. Second, melting is an example of second order phase change and consequently prone to hysteresis behaviour regardless of the utilized potential.  Last but not least of importance, here we consider a uniform oxide which cannot be met in an experimental study. Thus, we can rely on the qualitative behaviour and a quantitatively correct description is out of reach.

The velocity Verlet algorithm \citep{verlet1967,kateb2012} was utilized for time integration of the equation of motion with a timestep of 0.5~fs. We used Nose-Hoover thermostat with a damping of 3~fs. These conditions produce positions and velocities sampled from canonical (NVT) ensemble that utilized for the nanoparticles. For the nanowires the length is chosen equal to side length of simulation box with periodic boundary condition, to reduces the size effect along the nanowire axis. However, the box size in this direction must be changed to achieve zero average stress. To this end we considered the isothermal–isobaric ensemble (NPT) and applied zero pressure along the nanowire axis. The initial structure of partially oxidized Cu nanoparticle consists of a spherical Cu nanoparticle of 5~nm in diameter with a monolayer of Cu$_2$O ($\sim$0.25~nm) shell. The nanowire were considered to have the same diameter and oxide thickness with 20~nm length. It is worth mentioning that Cu$_2$O is the most stable form Cu oxide (cf.~\citet{tudela2008} and references therein). The initial velocities of the atoms were defined randomly from a Gaussian distribution corresponding to 5~K. After 10~ps relaxation, the system was heated to 50, 100, 150~K and so on and relaxed at each step using the desired ensemble and 2.5~ps, for both heating and relaxation steps. For the nanowire this heating causes a change in its length. 

The evolution of the system with the temperature is described by the total per atom potential energy $U_i^{\rm Tot}$. Values determined by averaging over constant temperature steps are indicated by $\langle\rangle_T$. The values with $\langle\rangle_{TN}$ are averaged over both constant temperature and atoms in the nanoparticle/nanowire. The OVITO package\footnote[2]{Version~3.0.0-dev481} was used for post-processing and producing atomistic illustrations. \citep{stukowski2009}


\section{Results and discussion}
\subsection{Caloric curves}
\begin{figure}[hbt!]
    \centering
    \includegraphics[width=.6\linewidth]{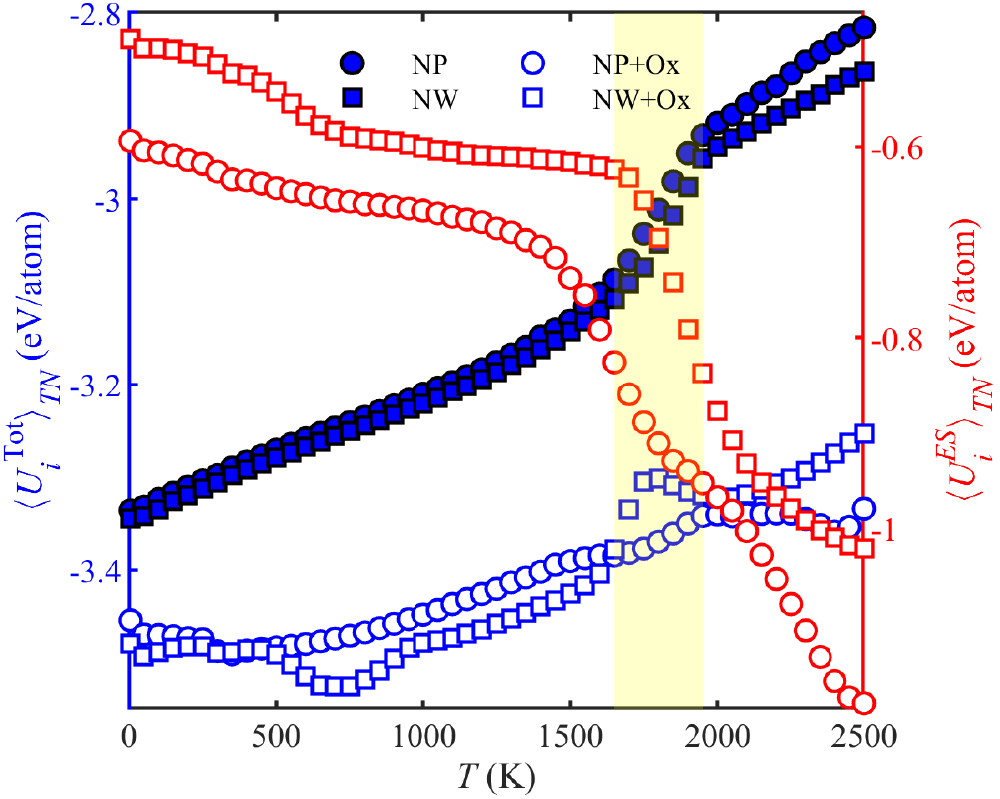}
    \caption{Variation of (blue) $\langle U_i^{\rm Tot}\rangle_{TN}$ and (red) $\langle U_i^{\rm ES}\rangle_{TN}$ with temperature $T$ for Cu nanoparticle (NP, circles) and nanowire (NW, squares) with (filled circles and squares) and without (empty circles and squares) oxide layer.}
    \label{fig:U}
\end{figure}

Since the release of latent heat of fusion ($\Delta H_{mn}$) causes an isotherm change in the so-called caloric curve(s), it is conventionally used to determine $T_{mn}$. The blue symbols and left axis in Fig.~\ref{fig:U} show  $\langle U_i^{\rm Tot}\rangle_{TN}$ variation with $T$ for nanoparticles and nanowires with and without oxide shell.  It can be seen that presence of surface oxide (\tikzcircle[blue]/\tikzsquare[blue]) causes a clear shift $\langle U_i^{\rm Tot}\rangle_{TN}$ towards lower energies. Cu particle (\tikzcircle[black, fill=blue]) and wire (\tikzsquare[black, fill=blue]) without surface oxide, present a linear variation with $T$ and a relatively isotherm transition in the 1650--1950~K (highlighted region). Such a transition is suppressed in the presence of the surface oxide on the nanoparticle (\tikzcircle[blue]). For the nanowire with surface oxide (\tikzsquare[blue]) the change occurs between 1600--1750~K. However it drops between 1800--2000~K that is not the case without surface oxide (\tikzsquare[black, fill=blue]). Also, the nanowire with surface oxide (\tikzsquare[blue]) presents a drop around 750~K which is characteristic of solid-state phase transition. \citep{kateb2018} 
Note that we did not observed such a transition for the nanowire without surface oxide (\tikzsquare[black, fill=blue]) in agreement with \citet{zhang2018}. They reported solid-state transition of pure Cu nanowires with diameter of 4.3~nm and smaller while diameters larger than 5~nm remained stable up to $T_{mn}$.

In order to understand the origin of the suppression, we plotted $\langle U_i^{\rm ES}\rangle_{TN}$ for the cases with surface oxide (\tikzcircle[red]/\tikzsquare[red]) on the right axis. It can be seen that $U_i^{\rm ES}$ presents a considerable drop for both cases. The drop for the nanoparticle (\tikzcircle[red]) starts at 1500~K which completely suppresses isotherm change in $\langle U_i^{\rm Tot}\rangle_{TN}$ (\tikzcircle[blue]). But for the nanowire (\tikzsquare[red]) $\langle U_i^{\rm ES}\rangle_{TN}$ drops at 1750~K and we can still observe isotherm change of $\langle U_i^{\rm Tot}\rangle_{TN}$ (\tikzsquare[blue]) up to this point. This, so-called \emph{electrostatic relaxation} will be discussed further in Subsection \ref{sec:Ox}.

\begin{figure}[hbt!]
    \centering
    \includegraphics[width=.265\linewidth]{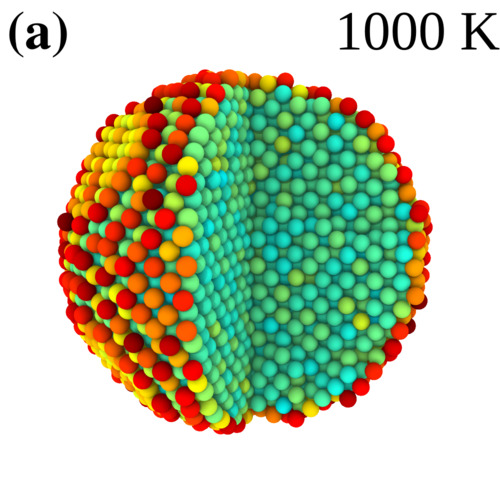}
    \includegraphics[width=.265\linewidth]{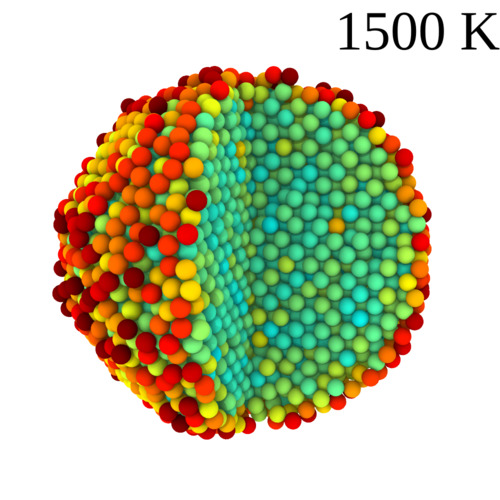}
    \includegraphics[width=.265\linewidth]{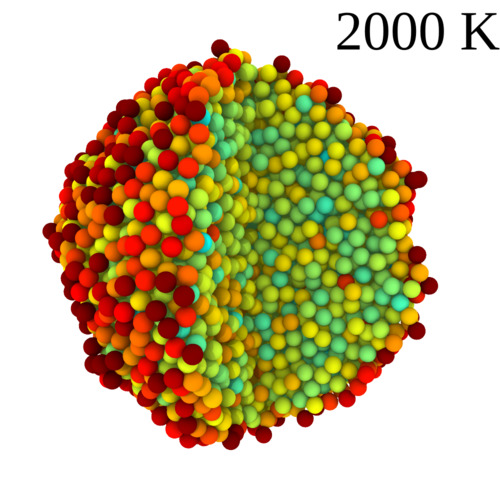}\\
    \includegraphics[width=.265\linewidth]{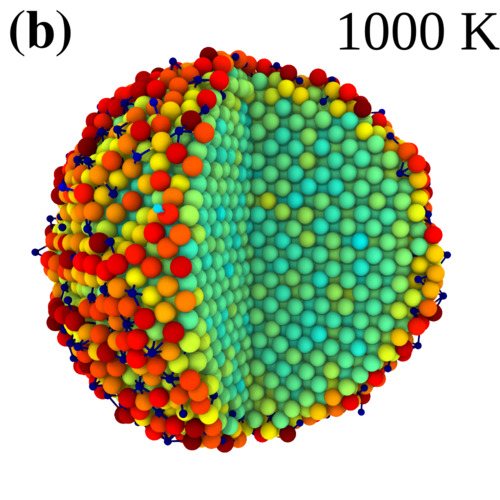}
    \includegraphics[width=.265\linewidth]{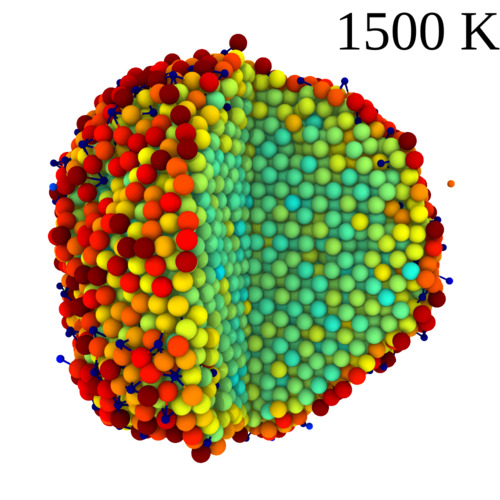}
    \includegraphics[width=.265\linewidth]{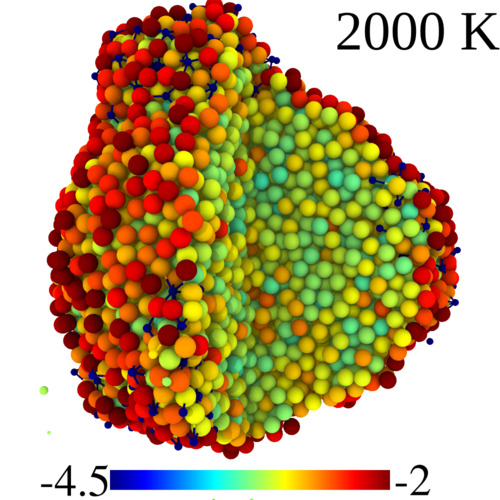}\\
    \includegraphics[width=.265\linewidth]{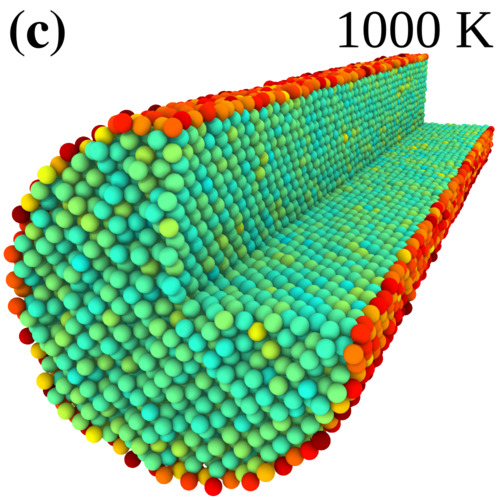}
    \includegraphics[width=.265\linewidth]{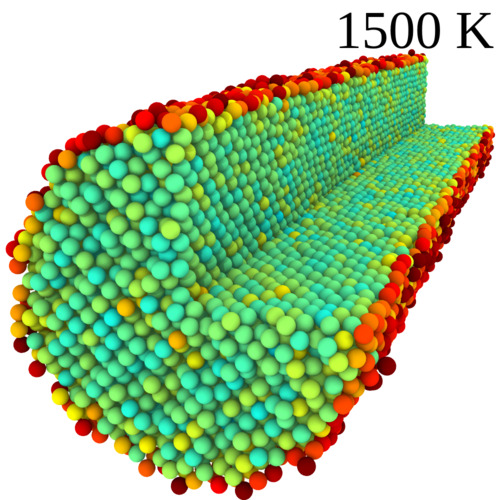}
    \includegraphics[width=.265\linewidth]{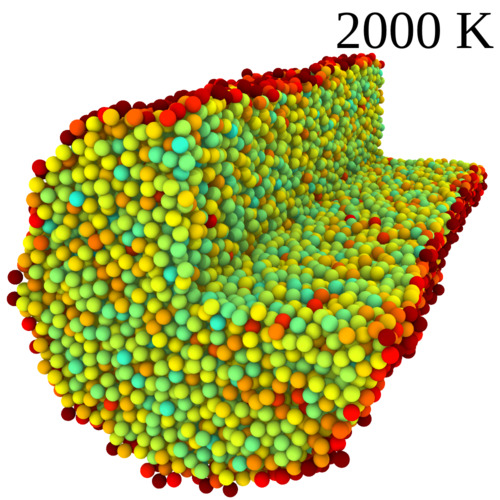}\\
    \includegraphics[width=.265\linewidth]{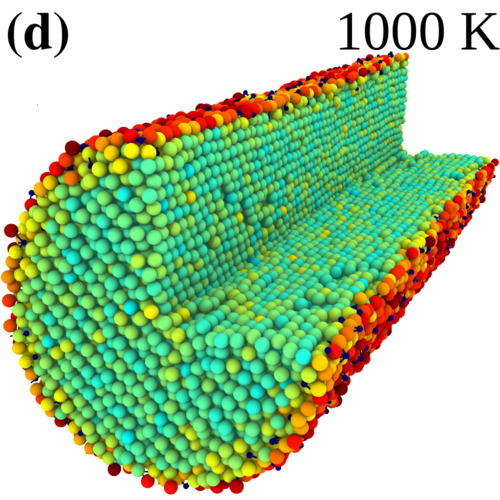}
    \includegraphics[width=.265\linewidth]{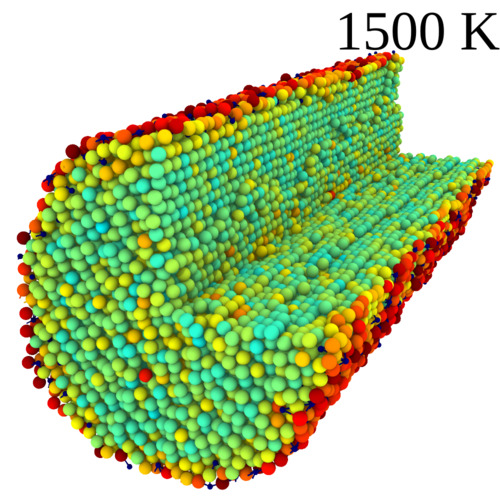}
    \includegraphics[width=.265\linewidth]{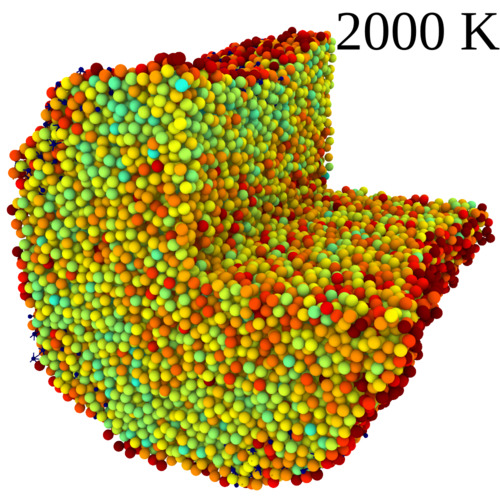}\\
    \caption{Snapshots of nanoparticles (a) without and (b) with oxide shell and nanowires (c) without and (d) with oxide shell. The oxygen atoms are shown with smaller diameter and appear in dark blue. The colorbar indicates $U_i^{\rm Tot}$ in eV. A quarter of nanoparticles and nanowires is removed for illustration of interior atoms.}
    \label{fig:Usnap}
\end{figure}

Fig.~\ref{fig:Usnap} Shows the snapshots of all cases at different temperature with colorbar being per-atom potential energy ($U_i^{\rm Tot}$) in eV. The latter value is averaged in Fig.~\ref{fig:U} and its distribution plotted in Fig.~\ref{fig:hist}. It can be seen that unlike 1000 and 1500~K snapshots, there are no atomic arrangement at 2000~K. Thus the suppression of isotherm change for the particle with surface oxide (\tikzcircle[blue] in Fig.~\ref{fig:U}) cannot be considered as melting resistance but deficiency of caloric curve to detect the melting. It is worth mentioning that Cu atoms, indicated by larger spheres, present a similar color distribution, meaning their $U_i^{\rm Tot}$ is the same. However, O atoms, indicated by smaller spheres in Fig.~\ref{fig:Usnap}(b) and (d), appear in dark blue. Thus, one can conclude that O atoms are responsible for lowering $\langle U_i^{\rm Tot}\rangle_{TN}$ compared to cases without the surface oxide i.e.\ \tikzcircle[blue]/\tikzsquare[blue] versus \tikzcircle[black, fill=blue]/\tikzsquare[black, fill=blue], respectively, in Fig.~\ref{fig:U}. However, lower $U_i$ of O atoms cannot explain complete suppression of isotherm change for the partially oxidized nanoparticle (\tikzcircle[blue]), or the drop after isotherm for the nanowire with surface oxide (\tikzsquare[blue]) observed in Fig.~\ref{fig:U}.
%

\subsection{Structure analysis \label{sec:struc}}
It is widely accepted that caloric curves must be compared with the structure analysis method to provide a comprehensive picture of a solid-state transition or melting. Various methods are utilized in atomistic simulations to identify the melting process based on the structure (cf.~\citet{kateb2018} and Refs.\ therein). Here we present and 
briefly discuss the result of a few structure analyses.

\subsubsection{Radial distribution function}
\begin{figure}[hbt!]
    \centering
    \includegraphics[width=.6\linewidth]{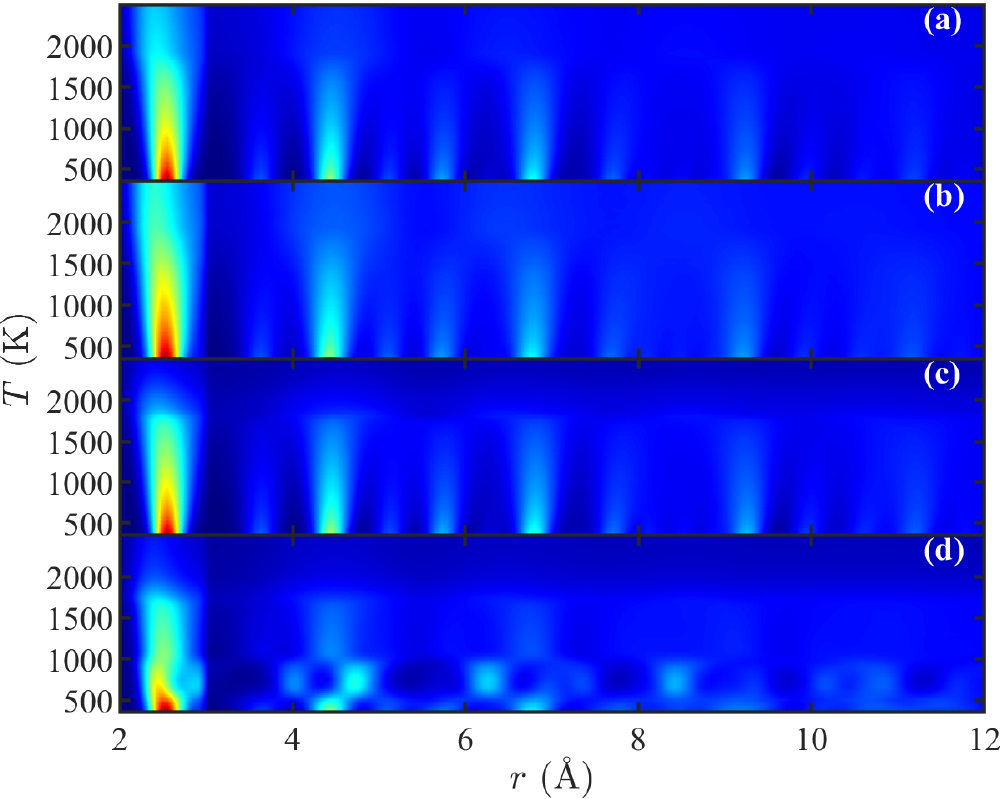}
    \caption{Variation of the normalized $\langle g(r_{\rm CuCu})\rangle_{TN}$ with $T$ for Cu nanoparticles (a) without and (b) with oxide shell and Cu nanowires (c) without and (d) with oxide shell. The colorbar indicates $\langle g(r_{\rm CuCu})\rangle_{TN}$ intensity.}
    \label{fig:grCu}
\end{figure}

The radial distribution function ($g(r)$), or pair correlation function, \citep{iida1988} is the most common structure analysis utilized in atomistic simulations. It gives atomic density variation with respect to the distance from a reference particle (atom here). The partial $g(r)$ were calculated using 200~bins and 12~{\AA} cutoff and then averaged over constant temperature steps. Fig.~\ref{fig:grCu} depicts the $\langle g(r_{\rm CuCu})\rangle_{TN}$ variation with $T$ and the colorbar indicates its intensity. It can be seen that the difference is insignificant for nanoparticles and for nanowires the major difference is the solid-state transition around 750~K. Also above 1700~K the major peak at 2.5~{\AA}, corresponding to the first nearest neighbors, vanishes for nanowires, Fig.~\ref{fig:grCu}(c) and (d), while for nanoparticles, (a) and (b), it shifts to smaller $r$. Thus, it appears that in the presence of the surface oxide $g(r)$ shows negligible difference and cannot be relied on for the case of nanoparticles. One can make a similar conclusion for the melting transition of nanowires.

\begin{figure}[hbt!]
    \centering
    \includegraphics[width=.6\linewidth]{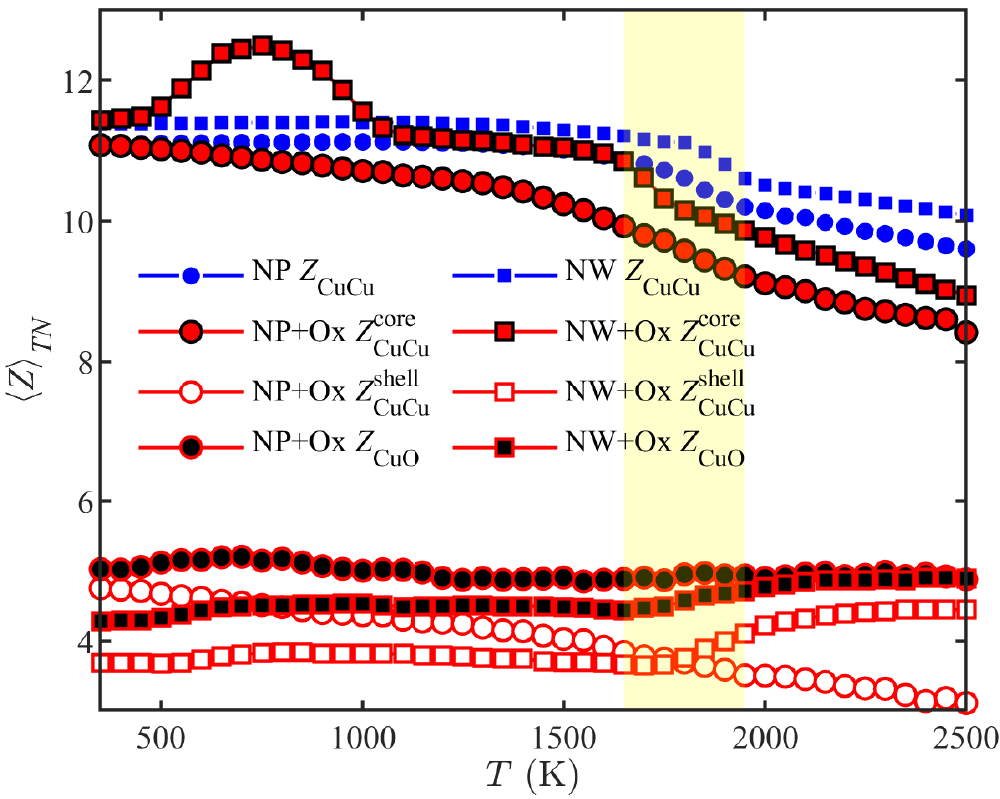}
\caption{Variation of $\langle Z\rangle_{T,N}$ with the $T$ for various pairs of first nearest neighbors. Highlighted region corresponds to isotherm changes of bare cases (cf.\ Fig.~\ref{fig:U})}
    \label{fig:cn}
\end{figure}

The integral of the major $\langle g(r)\rangle_{TN}$ peak gives the average coordination number ($\langle Z\rangle_{TN}$) or number of first nearest neighbors. It has a key role in modeling various properties such as $T_{mn}$, $\Delta H_{mn}$ \citep{safaei2008,attarian2008,safaei2010} and $\gamma_n$ \citep{jiang2008}. Fig.~\ref{fig:cn} shows the variation of $\langle Z\rangle_{TN}$ with $T$ for all cases studied here. For the cases with the surface oxide, different pairs of first nearest neighbors are provided those indicated by variant of red symbols. In the cases without the surface oxide, indicated by smaller filled symbols in blue (\tikzcircle[white, fill=blue]/\tikzsquare[white, fill=blue]), one can see a gradual change above 1500~K and a slop change at 1950~K which is more pronounced for the nanowire (\tikzsquare[white, fill=blue]). The later corresponds to the end of isotherm change in Fig.~\ref{fig:U}. Again, for the nanoparticle with the surface oxide $Z_{\rm CuCu}^{\rm core}$ (\tikzcircle[black, fill=red]) and $Z_{\rm CuCu}^{\rm shell}$ (\tikzcircle[red]) show very smooth changes. As expected $Z_{\rm CuCu}^{\rm shell}$ is much lower than that of core atoms. For the nanowire with oxide shell $Z_{\rm CuCu}^{\rm core}$ (\tikzsquare[black, fill=red]) a change around 750~K and a step change between 1600--1750~K were observed. However, its $Z_{\rm CuCu}^{\rm shell}$ (\tikzsquare[red]) shows an increase after melting (1750~K) which is also the case for its $Z_{\rm CuO}$ (\tikzsquare[red,fill=black]). Such an increase is associated with the shrinkage of oxide, cf.~Fig.~\ref{fig:q}(b). It is worth mentioning that, the oxide shrinkage also occur es for the nanoparticle, cf.~Fig.~\ref{fig:q}(b). However, its $Z_{\rm CuO}$ (\tikzcircle[red,fill=black]) does not increase due to the desorption of O.
Again, it is shown that utilizing $\langle Z\rangle_{TN}$ is tricky and may not be a proper melting criterion for the case of nanoparticle with surface oxide.

\subsubsection{Lindemann index}
Another widely used structure criterion introduced by \citet{lindemann1910} which accounts for the amplitude of vibrations rather than relative position of atoms
\begin{equation}
    \langle\delta_{\rm L}\rangle_{TN}=\frac{2}{N(N-1)}\sum_{i\neq j}\frac{\sqrt{\langle r_{ij}^2\rangle_T-\langle r_{ij}\rangle_T^2}}{\langle r_{ij}\rangle_T}
	\label{eq:delta}
\end{equation}
here $N$ is the total number of atoms in the system, $r_{ij}$ is the distance between particles $i$ and $j$. However, as pointed out by \citet{alavi2006}, and systematically studied by \citet{zhang2013}, it is difficult to observe a clear step change corresponding to the melting at the nanoscale. Such an issue originates from more degree of freedom at the surface. This is a more serious issue for polygonal nanoparticles where edge and corner atoms vibrate freely \citep{neyts2009} and utilizing $\delta_{\rm L}$ can lead to the misinterpretation of surface melting \citep{kateb2018}. Thus, under some conditions interpreting $\delta_{\rm L}$ is tricky which, to some extent, can be improved mainly by averaging over constant temperature or projecting per-atom $\delta_{\rm L}$ into geometry \citep{neyts2009}.

\begin{figure}[hbt!]
    \centering
    \includegraphics[width=.6\linewidth]{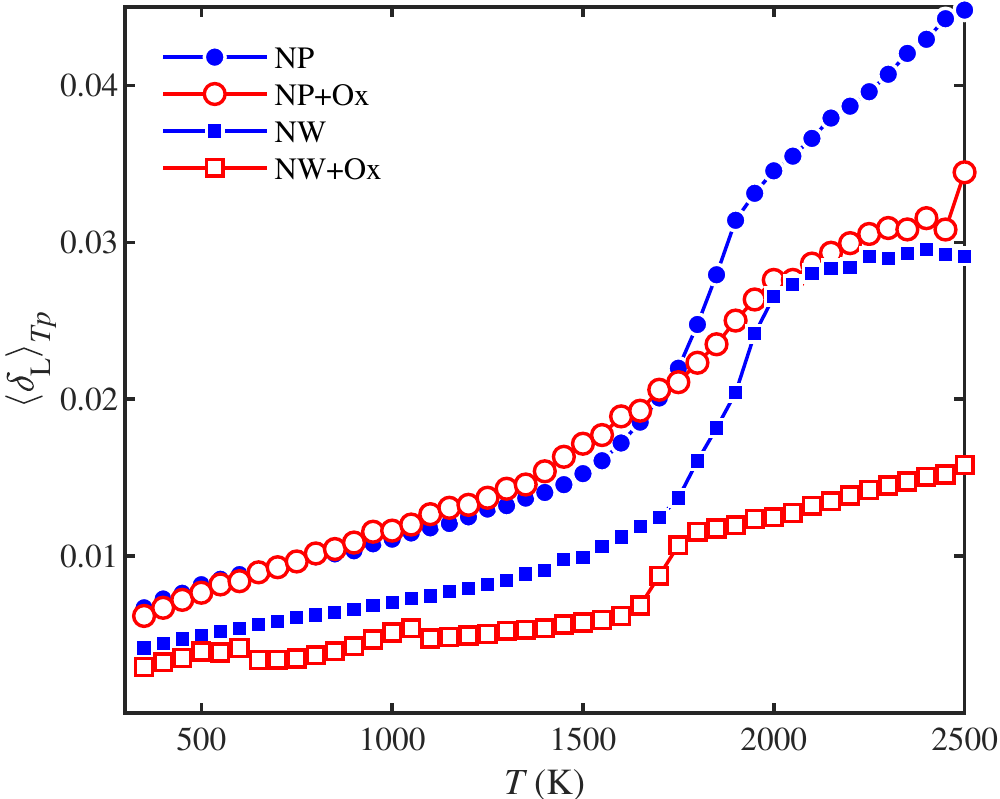}
    \caption{Variation of $\delta_{\rm L}$ with $T$ for Cu nanoparticle and nanowire with and without the oxide shell.}
    \label{fig:delta}
\end{figure}

Fig.~\ref{fig:delta} shows the variation of $\delta_{\rm L}$ with $T$ for both cases studied here. Here we escaped from the choice of polygonal shapes and averaged at constant temperature steps. These help detecting of isotherm step change more clearly for the cases without oxide (\tikzcircle[white, fill=blue]/\tikzsquare[white, fill=blue]). The presence of the oxide completely suppresses the step change for the nanoparticle (\tikzcircle[red]) and to some extent for the nanowire (\tikzsquare[red]). However, for the nanowire with surface oxide (\tikzsquare[red]) one can characterize three step changes, at 600 and 1050~K corresponding to the solid state phase transformation and around 1700~K due to melting. These temperatures are in agreement with the CNA results (cf.~Fig.~\ref{fig:cna}).  

\subsubsection{Common neighbor analysis}
Common neighbor analysis (CNA) is a more recent localized crystal structure characterization based on the relative position of two atoms that both are bonded to the so-called \textit{common neighbor} \citep{faken1994}. In particular it is sensitive to the symmetry of different pairs of bonded atoms and thus it can distinguish between fcc and hcp structures based on the (111) plane symmetry, i.e. central versus mirror symmetry, respectively. It is worth mentioning that OVITO enables utilizing an adaptive CNA \citep{stukowski2012,stukowski2014} that is not sensitive to the choice of cutoff for finding neighbors. However, CNA characterizes atoms located at the surface as disordered, due to the lack of symmetry \citep{kateb2019,kateb2020,kateb2021,kateb2020b} which can be an issue for the case of high surface to volume ratio.

\begin{figure}[hbt!]
    \centering
    \includegraphics[width=.6\linewidth]{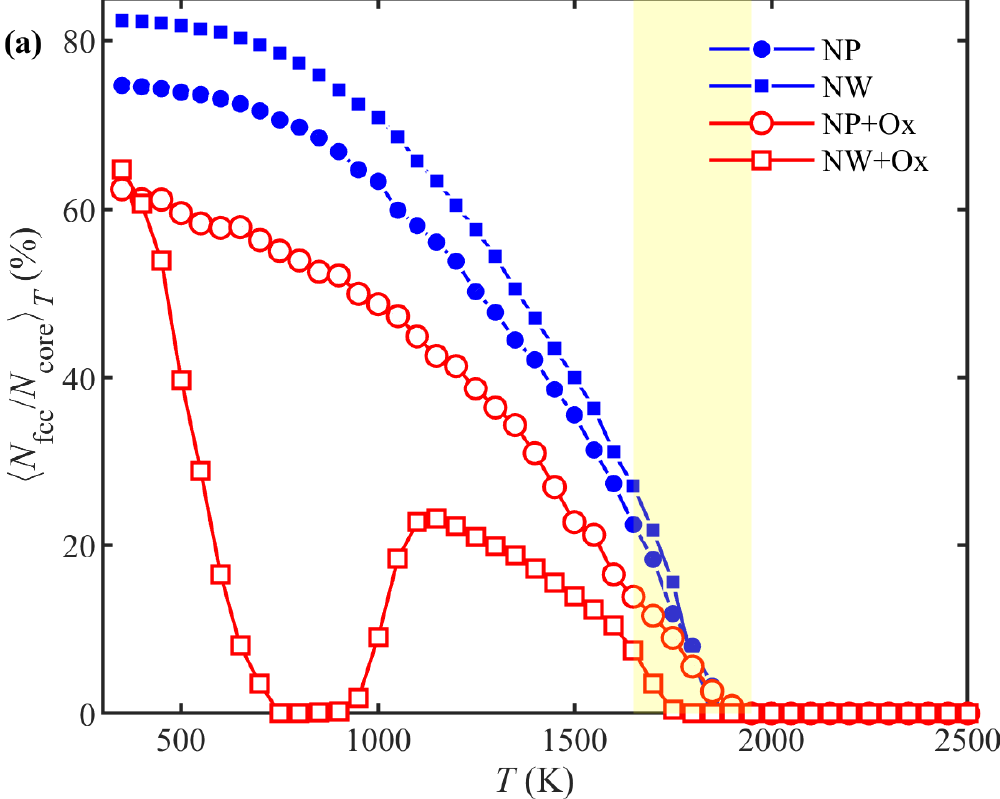}
	\\
	\includegraphics[width=.16\linewidth]{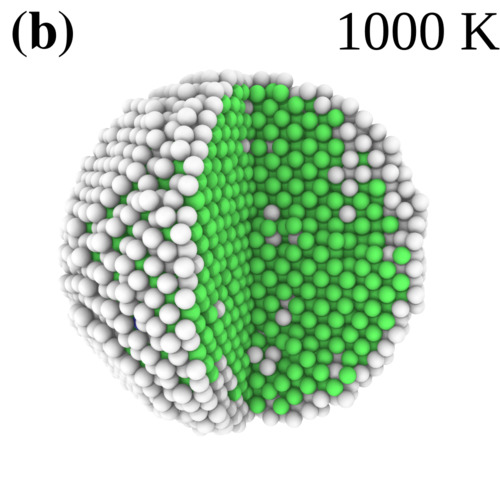}
	\includegraphics[width=.16\linewidth]{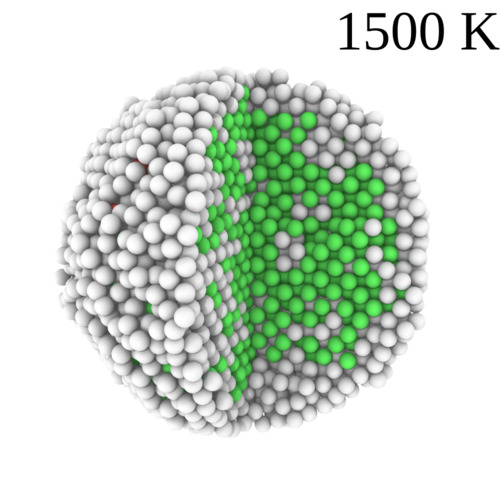}
	\includegraphics[width=.16\linewidth]{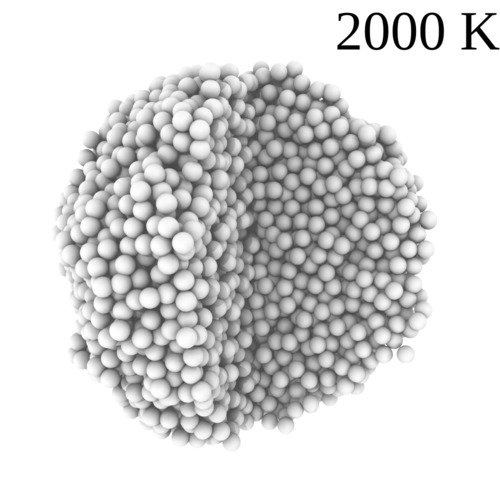}
	\includegraphics[width=.16\linewidth]{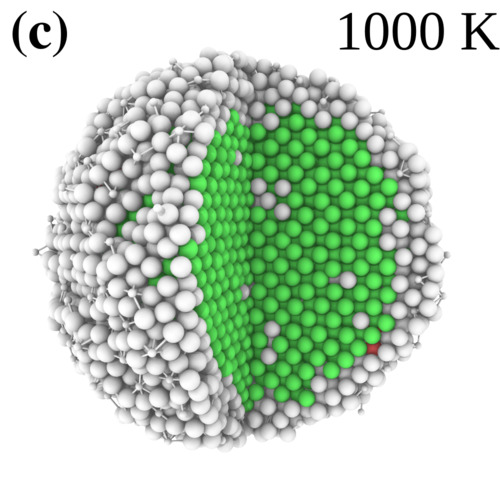}
	\includegraphics[width=.16\linewidth]{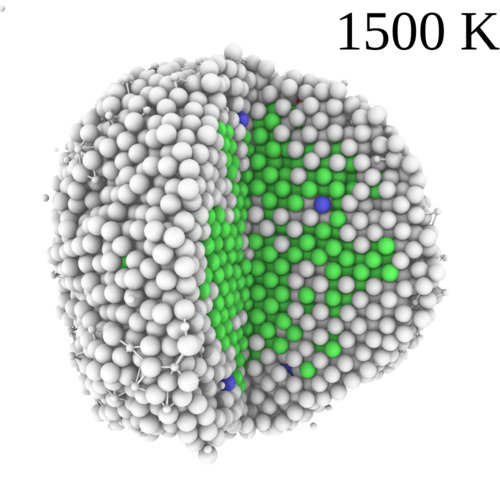}
	\includegraphics[width=.16\linewidth]{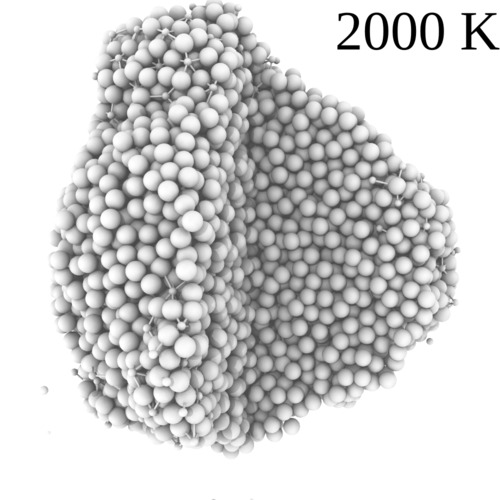}
	\\
	\includegraphics[width=.16\linewidth]{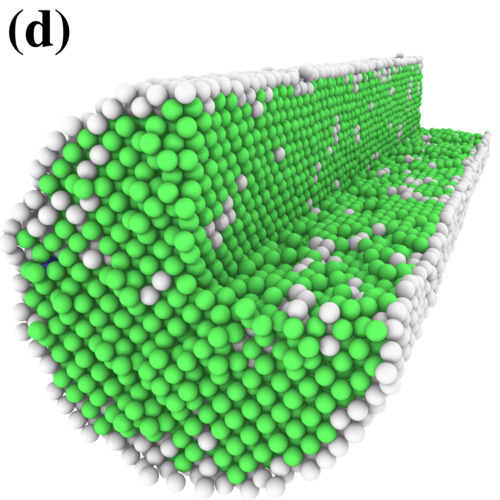}
	\includegraphics[width=.16\linewidth]{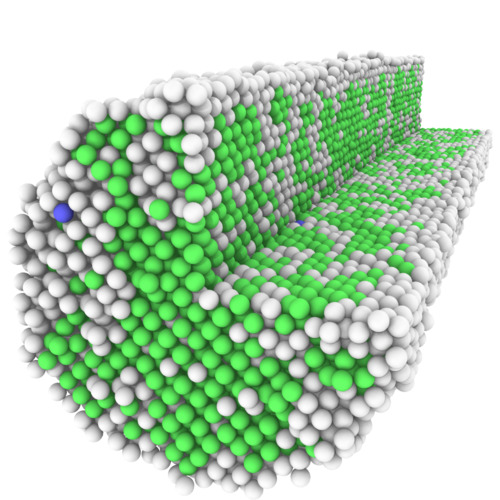}
	\includegraphics[width=.16\linewidth]{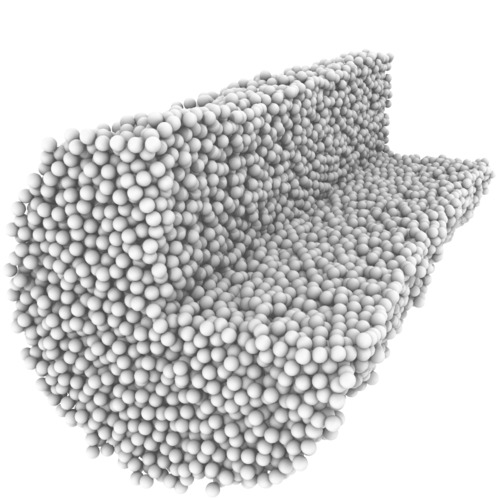}
	\includegraphics[width=.16\linewidth]{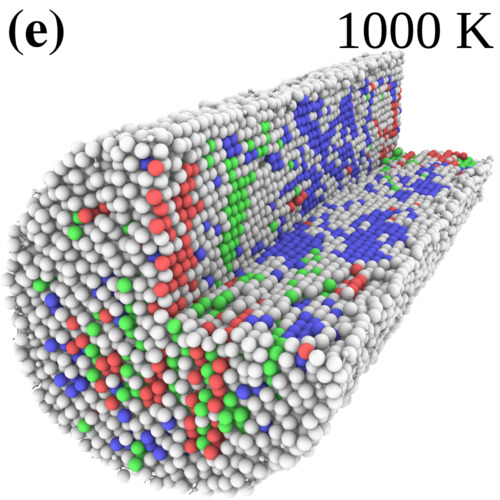}
	\includegraphics[width=.16\linewidth]{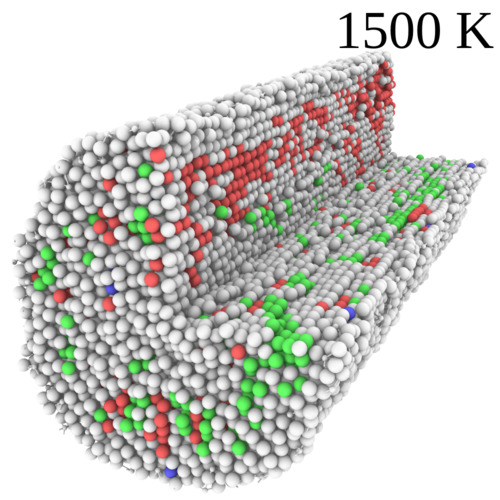}
	\includegraphics[width=.16\linewidth]{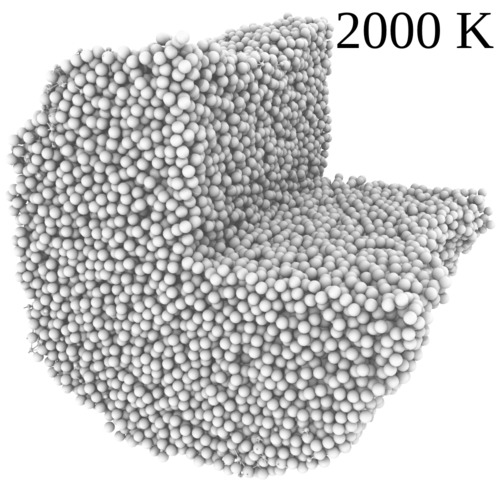}
    \caption{(a) Variation of fcc structure percentage with $T$ for the 5~nm Cu nanoparticle with and without 0.25~nm oxide shell. Corresponding snapshots of the nanoparticles (b) without and (c) with the oxide shell and nanowires (d) without and (e) with the oxide shell. The green, red, blue and white atoms indicate fcc, hcp, bcc and disordered structure, respectively.}
    \label{fig:cna}
\end{figure}

Fig.~\ref{fig:cna} presents the variation in the fcc percentage with $T$. Here Cu atoms in the core ($N_{\rm core}$) were only considered and oxide shell Cu atoms were excluded. It can bee seen that the surface oxide causes a noticeable drop in fcc atoms at 300~K, which is more pronounced in the nanowire (\tikzsquare[red]). Both cases without surface oxide (\tikzcircle[white, fill=blue]/\tikzsquare[white, fill=blue]) present no fcc atom at 1900~K which can be considered as complete melting. It appears that the surface oxide has a negligible effect on complete melting, 1950~K for the nanoparticle (\tikzcircle[red]) and 1800~K for the nanowire (\tikzsquare[red]). For the nanowire with surface oxide an phase transition occurs at low temperatures that gives no fcc atoms between 750--900~K. Comparison of nanowires (Fig.~\ref{fig:cna}(d) and (e)) at 1000 and 1500~K also shows that for the case with surface oxide, the ratio of fcc atoms (green) is decreased and there exist considerable number of bcc (blue), hcp (red) and disorderd (white) atoms in the core part. The later is an indication of phase transformation. 
%

\subsection{Potential energy distribution}
\begin{figure}[hbt!]
    \centering
    \includegraphics[width=.6\linewidth]{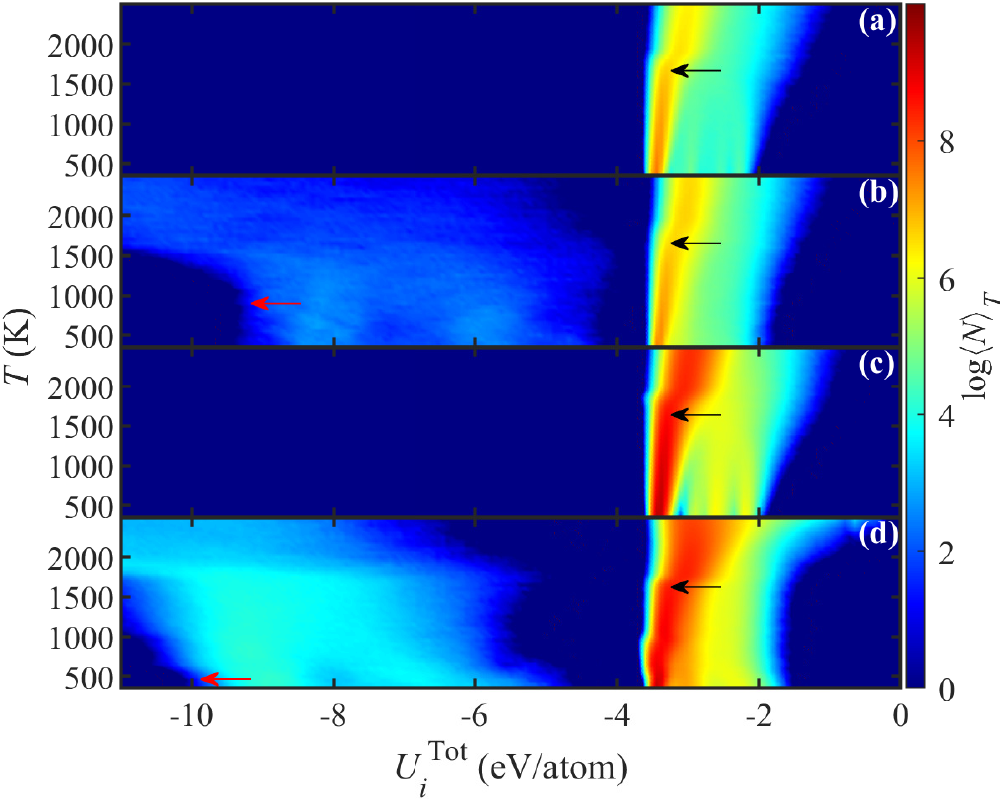}
    \caption{Variation of $\langle U_i^{\rm Tot}\rangle_T$ distribution with $T$ for Cu nanoparticles (a) without and (b) with surface oxide and nanowires (c) without and (d) with surface oxide. The colorbar indicates average counts (number of particles) at constant $T$ in log scale.}
    \label{fig:hist}
\end{figure}

We have recently shown that a quantity that brings caloric curves and structure analysis together is the distribution of $U_i^{\rm Tot}$. \citep{azadeh2019} This is mainly due to the fact that $U_i^{\rm Tot}$ is different for interior and surface atoms or for atoms on the surface of a polygonal nanoparticle it is different at the corner, edge and planes. By plotting $U_i^{\rm Tot}$ distribution one can maintain structure related features that are averaged in the caloric curves. Fig.~\ref{fig:hist} shows the $\langle U_i^{\rm Tot}\rangle_T$ distribution obtained by averaging over constant $T$ steps. The major peak in red belongs to interior atoms while more positive values belong to the surface atoms of the core Cu particle. The flat plateaus that appears as green in more negative ranges of Fig.~\ref{fig:hist}(b) and (d) belong to the oxygen atoms. The black arrows indicate the initial stage of the core melting which shows a more clear change than caloric curves (cf.~Fig.~\ref{fig:U}).

\subsection{Oxide shell dynamic \label{sec:Ox}}
\begin{figure}[hbt!]
    \centering
    \includegraphics[width=.265\linewidth]{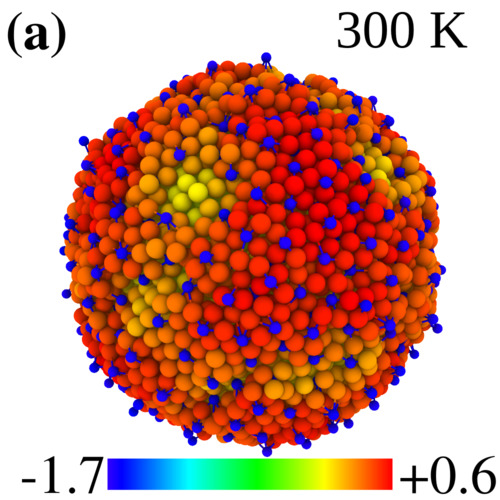}
	\includegraphics[width=.265\linewidth]{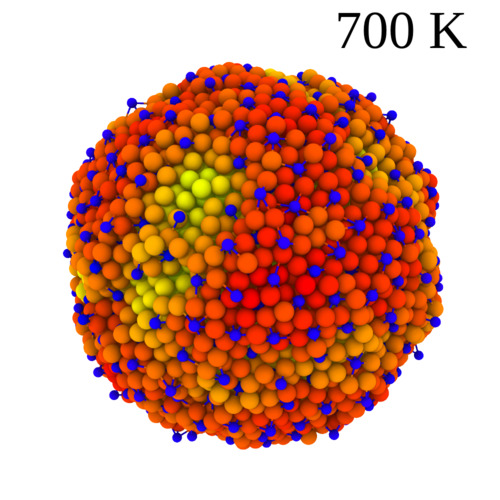}
	\includegraphics[width=.265\linewidth]{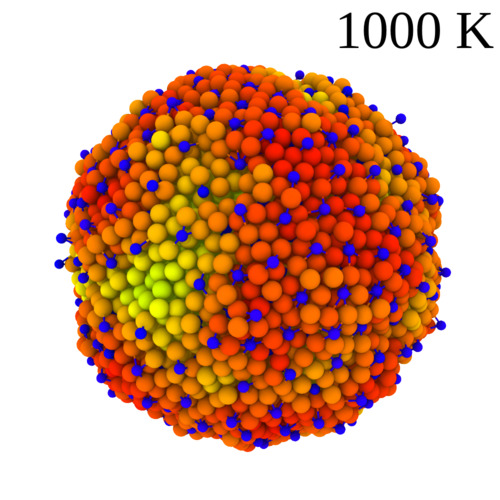}
	\\
	\includegraphics[width=.265\linewidth]{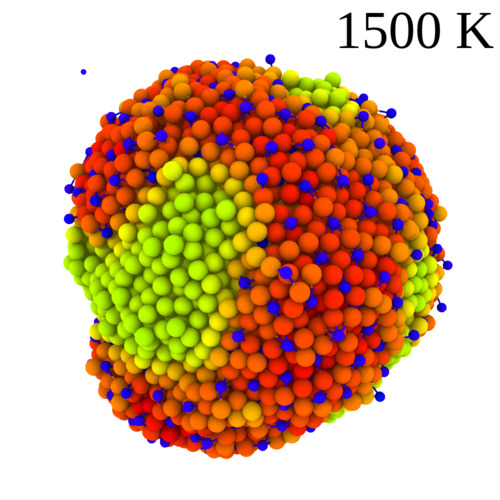}
	\includegraphics[width=.265\linewidth]{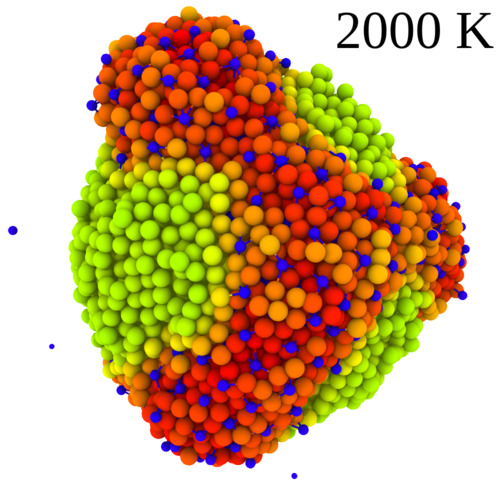}
	\includegraphics[width=.265\linewidth]{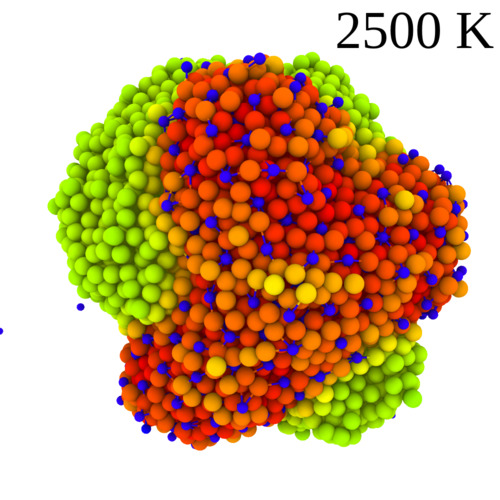}
	\\
	\includegraphics[width=.265\linewidth]{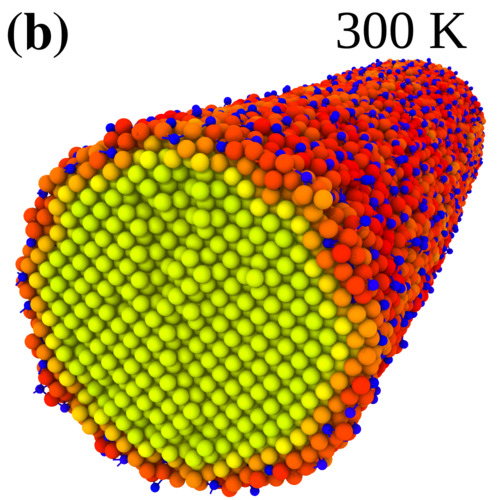}
	\includegraphics[width=.265\linewidth]{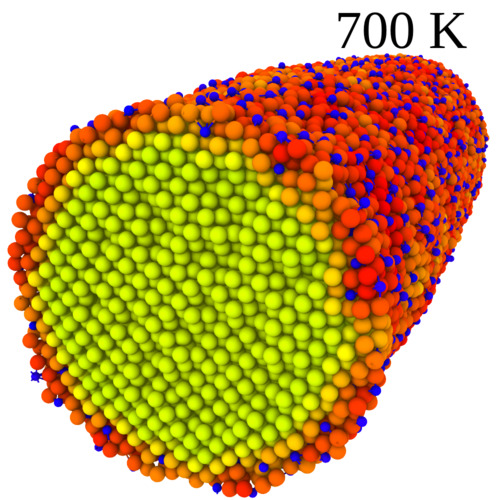}
	\includegraphics[width=.265\linewidth]{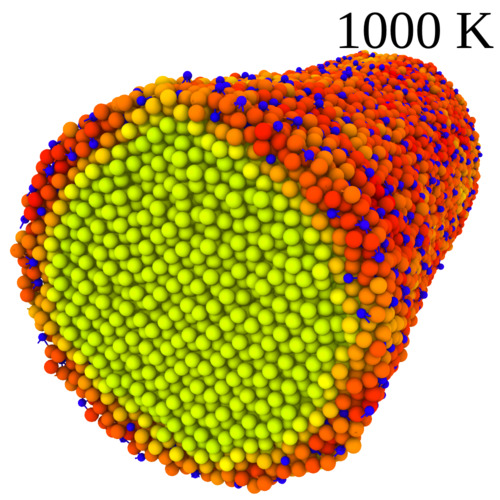}
	\\
	\includegraphics[width=.265\linewidth]{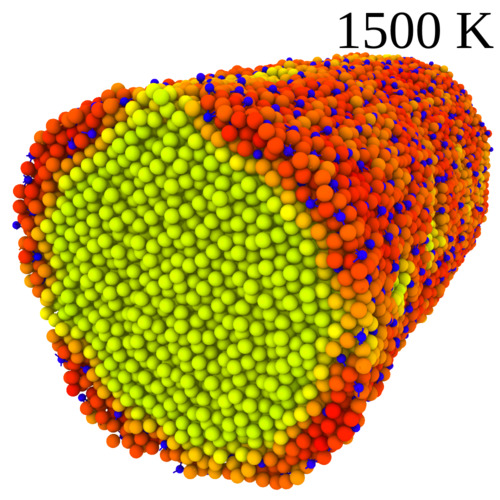}
	\includegraphics[width=.265\linewidth]{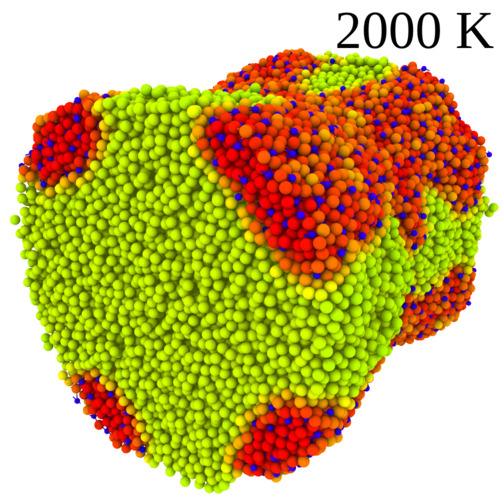}
	\includegraphics[width=.265\linewidth]{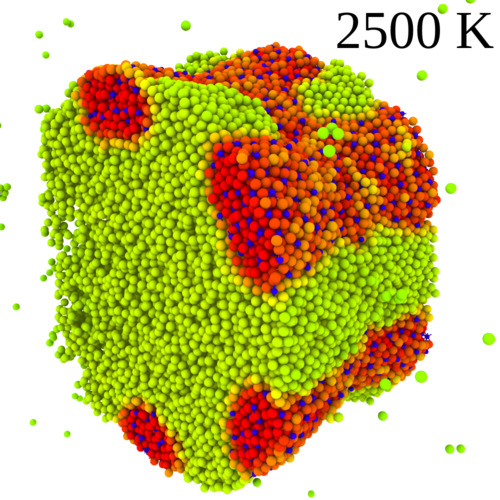}
    \caption{Snapshots of the oxide folding with $T$ for partially oxidized Cu (a) nanoparticle and (b) nanowire. The colorbar indicates $q_i$ with more oxidized atoms in red and metallic Cu in green. Oxygen atoms are shown with smaller spheres in deep blue.}
    \label{fig:q}
\end{figure}

Fig.~\ref{fig:q} shows the snapshots of the particle and nanowire with the surface oxide. The color bar indicates the partial electrostatic charge of the atom $i$, $q_i$, as an indicator of oxidation state i.e.\ red atoms being oxidized Cu, green ones are metallic and blue atoms are oxygen. At 297~K, $\sim$42\% of O atoms were found within the core radius for both nanoparticle and nanowire. This is in agreement with Cu oxidation experiments. \citep{leitner2020} Note that in general the time required for illustration of diffusion is several orders of magnitude longer than what can be achieved within MD simulation. Thus, we were only able to locate O atoms in vicinity of metal-oxide interface. It can be seen that for both cases oxide layers become thin until metallic Cu (in green) appears at the surface. Although thinning step starts at 297~K for the nanoparticle, the oxide is stable up to 1000~K. Then the oxides shrinks and becomes thicker which is more evident above 1500~K. The early stage of the shrinkage is indicated by red arrows in Fig.~\ref{fig:hist}(b) and (d). Unlike \citet{puri2010} we did not observed evaporation of oxide clusters. But for the case of nanoparticle, we observed desorption of oxygen atoms ($\sim$6\%) in between 1050--2500~K which accelerates in 1200--1650~K range. Besides the desorption only occurs at the apparent interface of metal-oxide. While for the nanoparticle no evaporation/sublimation of Cu atoms is observed, for the nanowire with the surface oxide Cu evaporation occurs above $\sim$2200~K.

It is observed that the oxide cluster is made of Cu atoms of both initial core and shell. The remaining oxide has Cu:O ratio of 4.17:1 suggesting the Cu$_4$O stoichiometry. Cu$_4$O is discovered by \citet{guan1984} as a meta-stable phase that may decompose to more stable Cu$_2$O and Cu. Later, tight bonding simulation showed that it is stable \citep{liu1999} at elevated temperatures. The shrinkage of oxide and consumption of metallic Cu for the formation Cu$_4$O are of practical interest regarding the conductive ink in printed circuits.

\section{Conclusions}
In conclusion, the stability and melting behavior of Cu nanoparticle and nanowire with or without a monolayer of Cu$_2$O shell is studied by MD simulation and COMB potential. Several melting criterion were utilized to provide a comprehensive picture of melting. In the presence of the oxide at the surface we observed a different melting behaviour than in the pure metal. This includes inwards diffusion of atomic oxygen into subsurface layers and shrinkage of oxide shell for both nanosolids. The latter is accompanied with formation of meta-stable Cu$_4$O at elevated temperatures by consuming some of the metallic Cu. We also observed a solid-state transition in nanowire induced by presence of oxide shell.

\begin{acknowledgement}
This work was supported by the Icelandic Research Fund, Grant No.~195943.
\end{acknowledgement}


\bibliography{Ref}

\end{singlespace}

\end{document}